\documentclass[%
reprint,
showpacs,
 amsmath,amssymb,
prb,
]{revtex4-2}

\usepackage{graphicx}
\usepackage{dcolumn}
\usepackage{bm}
\usepackage{pythonhighlight}
\usepackage{moresize}
\usepackage{hyperref}

\begin{document}
\title{Pyrovskite: A software package for the high throughput construction, analysis, and featurization of two- and three-dimensional perovskite systems}

\author{Robert Stanton}
\author{Dhara J. Trivedi}%
 \email{dtrivedi@clarkson.edu}
\affiliation{%
 Department of Physics, Clarkson University, Potsdam, NY 13699, USA
}%

\date{\today}

\begin{abstract}
The increased computational and experimental interest in perovskite systems comprising novel phases and reduced dimensionality has greatly expanded the search space for this class of materials. In similar fields, unified frameworks exist for the procedural generation and subsequent analysis of these complex condensed matter systems. Given the relatively recent rise in popularity of these novel perovskite phases, such a framework has yet to be created. In this work, we introduce Pyrovskite, an open source software package to aid in both the high-throughput and fine-grained generation, simulation, and subsequent analysis of this expanded family of perovskite systems. Additionally we introduce a new descriptor for octahedral distortions in systems including but not limited to perovskites. This descriptor quantifies diagonal displacements of the B-site cation in a BX$_6$ octahedral coordination environment, which has been shown to contribute to increased Rashba-Dresselhaus splitting in perovskite systems.
\end{abstract}

\maketitle

\section{\label{sec:level1}Introduction}

In the last decade, perovskite systems have become ubiquitous in the search for high efficiency photovoltaic devices \cite{wang2019review, zhang2022origin, correa2017rapid, hsiao2015fundamental}. With more recent developments, two-dimensional perovskites (2DPKs) have emerged as candidates which maintain or improve upon many advantageous properties of their bulk counterparts, such as low exciton binding energies, geometrically tunable electronic properties, and inexpensive building components, while addressing the key concerns of bulk perovskites which are long-term stability and device longevity \cite{krishna2019mixed, wu20212d, blancon2020semiconductor, stanton2022probing, vasileiadou2021shedding}. Additionally, 2DPKs serve as a platform for the realization of physical properties seldom seen in their bulk counterparts which make them favorable for usage beyond photovoltaics, such as Rashba-Dresselhaus splitting for spintronic applications, and strong electron-phonon coupling while maintaining narrow linewidths for light emitting applications \cite{steele2019role, li2023effect, leng2021ferroelectricity, bhattacharya2023spin}. The increased area of applicability, and thereby target properties, together with the widely expanded family of perovskites which serve as potential candidate structures for materials scientists poses both experimental and computational challenges. Large-scale, data driven, computational approaches in the realm of hybrid perovskites which contain organic cations remain limited in the literature, particularly in the realm of 2DPKs \cite{PhysRevB.107.174109, zhao2022high, marchenko2020database, kim2017hybrid, HU2023105841}. This is in part due to the complexity of the resultant structure, as well as the size of simulation cells which frequently enter into the hundreds of atoms for even modest layer thicknesses.\\

Another challenge in dealing with perovskite systems is the structural diversity associated with the family of materials. Bulk ABX$_{3}$ perovskites, double A$_{2}$BB’X$_{6}$ perovskites, 2DPKs and 2D-double perovskites form an extremely diverse set of materials in terms of phase, chemical composition, and their resulting geometric properties \cite{wolf2021doubling, guo2021progress, quan2022two}. With this being the case, structure generation and characterization becomes a challenge for which a comprehensive suite of tools, robust to the varied perovskite phases, dimensionalities, and diversity of organic cations has yet to be created to the author’s knowledge. Characterizing the geometric properties associated with the relaxed perovskite crystal structures is of vital importance, as numerous structural features have been established as effective means of tuning the electronic properties of the resulting material. For example, the literature is abundant with investigations characterizing the impact of organic cation orientation, octahedral tilting, organic spacer selection in 2D-perovskites, and octahedral distortions on the resulting electronic structure associated with the perovskite system \cite{yin2023tuning, krach2023emergence, ji2023jahn, stanton2022atomistic, li2016atomic}.\\

In this work, we introduce a software package, Pyrovskite, as a comprehensive suite of utilities for the 1) generation of perovskite crystal structures and electronic structure input files, 2) featurization of crystal structures with octahedral distortion parameters, 3) production of publication quality figures concerning the geometric properties of the system, and 4) interfacing the abovementioned information with typical data structures used for subsequent data analysis or machine learning. Additionally, motivated by the work of Maurer et al. regarding the link between diagonal B-site cation displacements and increased Rashba-Dresselhaus spin-splitting, we introduce a new descriptor quantifying these diagonal displacements, and compare it with pre-existing octahedral distortion parameters \cite{maurer2022rashba}.\\

The paper is structured as follows: In Sec.~\ref{sec:level2}, we discuss the relevance of distortions to the octahedral coordination environments present in many systems including perovskites, and we review the ubiquitous descriptors in the literature which quantify such distortions. Additionally, we highlight the difficulty, and thereby lack of computational solutions, associated with procedural generation of the expanded family of perovskites discussed in the introduction. In Sec.~\ref{sec:level3}, we introduce a new descriptor for octahedral distortions which can be computed with the Pyrovskite package to quantify diagonal displacements of the B-site cation in BX$_{6}$ octahedral coordination environments including but not limited to perovskites. Additionally, we discuss the implementation of procedural perovskite generation present within Pyrovskite which is capable of addressing inorganic, hybrid, and double perovskites of varying phase and dimensionality. In Sec.~\ref{sec:level4}, we comprehensively illustrate the capabilities of the package, including structure generation and characterization, the computation of geometric properties such as octahedral distortions, and the available interface with electronic structure and data analysis packages for machine learning. Finally, we present Sec.~\ref{sec:level5} containing conclusions and an outlook for future improvements to the package and its area of applicability.\\

\section{\label{sec:level2}Problem Formulation}

\subsection{\label{sec:level21}Descriptors for distortions in octahedral coordination environments}

Octahedral coordination environments are present in a myriad of systems relevant to modern materials science, from spin-crossover and Werner-type complexes to bulk metal-oxide systems and of particular interest to the present work, perovskites \cite{jang2021understanding, kim2006missing, su2022direct, zhou2022two, zhou2023octahedral, hogue2018spin, wei2022polarization}. Common amongst these materials are distortions of the idealized octahedral coordination environment in the form of deviations in bond lengths and angles away from those present in the perfectly symmetric octahedra. These distortions are typically facilitated by the geometry or electronic structure of the local chemical environment, for example by organic A-site cations in hybrid perovskites or Jan-Teller distortions in the case of transition-metal complexes \cite{wang2019atomistic, xiao2020ultrafast, liu2022effects}. The impact of the abovementioned distortions on the resulting material properties have been the subject of considerable investigation for decades \cite{radaelli1996charge, he2010control, zhou2005universal}. Given that these octahedral distortions typically serve as tuning mechanisms for a target material property, such as the band gap in a condensed matter system, a number of descriptors exist dating back to 1971 by Robinson $\textit{et al.}$ to distinguish and quantify different types of deviations from the idealized octahedral environment \cite{huang2016electronic, robinson1971quadratic}. Two such descriptors describing the deviations are $\Delta$ and $\Sigma$ representing deviations in bond lengths and angles respectively, given by Eq.~\ref{eq:delta} and~\ref{eq:sigma}.

\begin{equation}
\Delta = \frac{1}{6}\sum_{i=1}^{6}\left(\frac{d_{i}-d}{d}\right)^{2}
\label{eq:delta}
\end{equation}

\begin{equation}
\Sigma = \frac{1}{6}\sum_{i=1}^{12}\mid\phi_{i}-90^\circ\mid
\label{eq:sigma}
\end{equation}

In the context of BX$_{6}$ octahedra, $d_{i}$ is the distance from B-cation to the $i'$th X-anion, $d$ is the average $B-X$ distance, and $\phi_{i}$ is the $i'$th \textit{cis} $X-B-X$ angle in the octahedra. The notation as presented is adopted from by Ketkaew $\textit{et al.}$, and the quantities are in essence the same as those described by Robinson $\textit{et al.}$ \cite{D0DT03988H, robinson1971quadratic}. An additional descriptor was formulated by Ketkaew $\textit{et al.}$ quantifying twisting of the triangular octahedral faces, however for this feature we direct the reader to the OctaDist codebase \cite{D0DT03988H}.

A recently highlighted form of octahedral distortion comes with the dislocation of B-site cations found in the BX$_{6}$ octahedra present in perovskite systems. The B-site displacement has been demonstrated by Maurer $\textit{et al.}$ to be tied to the strength of Rashba-Dresselhaus spin-splitting exhibited by the perovskite system which is of crucial importance to prospective spintronic applications \cite{maurer2022rashba}. In particular, the Rashba spin-splitting was demonstrated to correlate with diagonal displacements of the B-site cation relative to the octahedral coordination environment. In Sec.~\ref{sec:level3}, we formulate a descriptor to quantify diagonal displacements of the B-site cations, and discuss their implementation in the Pyrovskite package.

\subsection{\label{sec:level22}Procedural generation of hybrid perovskite systems}

The increasing relevance of double perovskites, hybrid perovskites, 2DPKs, and their combinations for photovoltaic, thermoelectric, ultraviolet radiation detection, and light-emitting applications has greatly increased the materials search space relative to the previous consideration of only 3D perovskites \cite{sajjad2020ultralow, aslam2021structural, xu20203d, zhang2021high, chakraborty2022quantum}. The 2DPKs in particular contribute greatly to this increased structural diversity, as the phase dependent chemical formula of A’$_{2}$A$_{n-1}$BX$_{3n+1}$ for the 2D Ruddlesden-Popper (2DRP) and monolayer configurations, and A’$_{2}$A$_{n-1}$BX$_{3n+1}$ for the Dion-Jacobson (DJ) phase introduce two degrees of freedom not present in bulk perovskites. The A’ spacing cation serves to decouple the adjacent inorganic networks of BX$_{6}$ octahedra both spatially, and electronically \cite{filip2022screening}. The choice of spacing cation has been demonstrated to have non-trivial effects on properties ranging from the structural stability, device longevity, and even electronic structure of the resulting material \cite{xu2022cspbi3, zheng2019two, yan2019benefiting}. Additionally, the layer thickness $n$ determines the number of inorganic octahedra per layer, with electronic band gaps converging from above to their bulk counterpart in the case of $n \to \infty$. These additional degrees of freedom create an entire family of structures for a given bulk ABX$_{3}$ perovskite, or A$_{2}$BB’X$_{6}$ double perovskite which often host very different physical properties in their dimensionally reduced form \cite{lei20222d}.

This large search space creates the immediately apparent problem for computational researchers of resource limitations. However, a more conspicuous problem is in the reliable crystal structure generation of perovskites containing organic A- and A'-site cations of varied length, size, and shape in the 2DPKs. Interlayer alignment for phase specificity, interlayer spacing dictated by the interactions between BX$_{6}$ octahedra and A’ cations, and the placement as well as orientation of organic A and A’ building components make the procedural generation of the newly diversified family of perovskite systems a challenge.

\section{\label{sec:level3}Implementation}

Herein, we introduce Pyrovskite, a Python package facilitating the high-throughput creation, simulation, and analysis of perovskite systems of varied phase, dimensionality, and chemical composition. Additionally, we note that all functionality associated with computation of octahedral distortions is applicable to any condensed phase or molecular system containing octahedral coordination environments. We first formalize the new $\Lambda_2$ and $\Lambda_3$ descriptors for quantifying the diagonal displacement of B-cations in BX$_6$ coordination environments. This is followed by a comprehensive overview of the user-friendly procedural perovskite generation present in the Pyrovskite package for bulk, double, 2D, and 2D-double perovskite phases.

\subsection{\label{sec:level31}Descriptor for the diagonal displacement of B-site cation}

\begin{figure}[!ht]
   \begin{centering}
    \includegraphics[width=0.9\columnwidth]{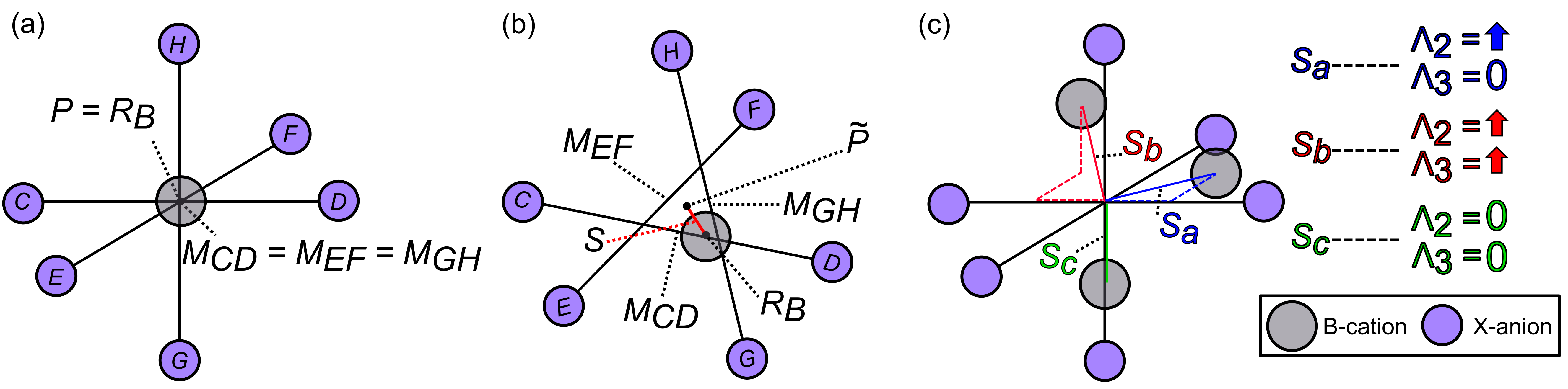}
      \caption{Schematic representation of an (a) undistorted, and (b) distorted BX$_{6}$ octahedra with the quantities from Sec.~\ref{sec:level31} relevant to the definition of the $\Lambda_2$ and $\Lambda_3$ descriptors labelled. Distortions in (b) and (c) are exaggerated relative to typical octahedral distortions for clarity. (c) Three example placements of the B-cation in the BX$_{6}$ octahedra and the subsequent qualitative effect on the $\Lambda_2$ and $\Lambda_3$ descriptors.}
      \label{fig:octahedra}
   \end{centering}
\end{figure}

In a BX$_{6}$ octahedra respecting full octahedral symmetry, the midpoint of the three lines connecting $\textit{trans}$-pairs of X atoms, $M_{CD}$, $M_{EF}$, and $M_{GH}$ coincide directly with one another, and with the B-site cation at $R_{B}$ in the center of the octahedra, termed $P$ [as depicted in Fig.~\ref{fig:octahedra}(a)]. 

\begin{equation}
M_{CD}=M_{EF}=M_{GH}=P=R_{B}
\label{eq:pristine}
\end{equation}

Upon octahedral distortion, however the midpoints of $\textit{trans}$-X-X connections separate, and the natural center of the octahedra becomes poorly defined. To remediate this problem, we define the natural center $\tilde{P}$ of the BX$_{6}$ octahedra by considering the center of the three midpoints of the $\textit{trans}$-pairs of X atoms, as in Eq.~\ref{eq:ptilde}:

\begin{equation}
\tilde{P}=\frac{1}{3}(M_{CD}+M_{EF}+M_{GH})
\label{eq:ptilde}
\end{equation}

The displacement vector $S$ between the B-cation and the natural center is given by Eq.~\ref{eq:svec}:

\begin{equation}
S=R_{B}-\tilde{P}
\label{eq:svec}
\end{equation}

$\tilde{P}$ and $S$ can be seen in the distorted octahedra in Fig.~\ref{fig:octahedra}(b). In order to characterize the diagonal nature of the displacement vector $S$, we require reference to a natural basis of the given octahedra. To achieve this, we form the non-orthogonal basis (Eq.~\ref{eq:unitvec}) as the collection of unit vectors associated with the lines connecting the three $\textit{trans}$-pairs of X atoms, $CD$, $EF$, and $GH$.

\begin{equation}
\{e_{1},e_{2},e_{3}\}=\left\{\frac{CD}{|CD|}, \frac{EF}{|EF|}, \frac{GH}{|GH|}\right\}
\label{eq:unitvec}
\end{equation}

where, $CD=R_{C}-R_{D}$ and so on. The characterization of a diagonal displacement of the B-site cation is then informed by the ratio of the components of $S$ when projected onto the basis of the octahedra. We define $\lambda'_{ij}$ for the three unique pairs of $S$ components in Eq.~\ref{eq:lambda} such that $\lambda'_{ij}$ are bounded by $0$ and $1$. A $\lambda'_{ij}$ value approaching 0 then corresponds to a disproportionate displacement almost entirely along the $i$, or $j$ direction, and a $\lambda'_{ij}$ value approaching 1 corresponds to nearly equal displacement along the $i$ and $j$ directions. To differentiate between equally diagonal displacements (that is, collinear displacements relative to the natural center of the octahedra $\tilde{P}$) of different magnitude, the $\lambda_{ij}$ values are then obtained by scaling $\lambda'_{ij}$ by a factor of $(|S_i |+|S_j |)$.

\begin{equation}
\lambda'_{ij}=\textsf{min}\left(\frac{|S_{i}|}{|S_{j}|}, \frac{|S_{j}|}{|S_{i}|}\right)
\label{eq:lambda}
\end{equation}

We then define a new set of descriptors which indicate diagonal displacements within one plane spanned by a pair of basis vectors of the octahedra, and along all three planes spanned by the pairs of basis vectors as $\Lambda_{2}$ and $\Lambda_{3}$ respectively. Fig.~\ref{fig:octahedra}(c) gives three examples of B-cation displacements and the qualitative corresponding impact on the $\Lambda_2$ and $\Lambda_3$ descriptors.\\

\begin{eqnarray}
\Lambda_{2} &=&\textsf{max}(\lambda_{12},\lambda_{13},\lambda_{23})\\
\Lambda_{3}&=&\lambda_{12}\lambda_{13}\lambda_{23}
\end{eqnarray}

\begin{figure}[!ht]
   \begin{centering}
    \includegraphics[width=0.9\columnwidth]{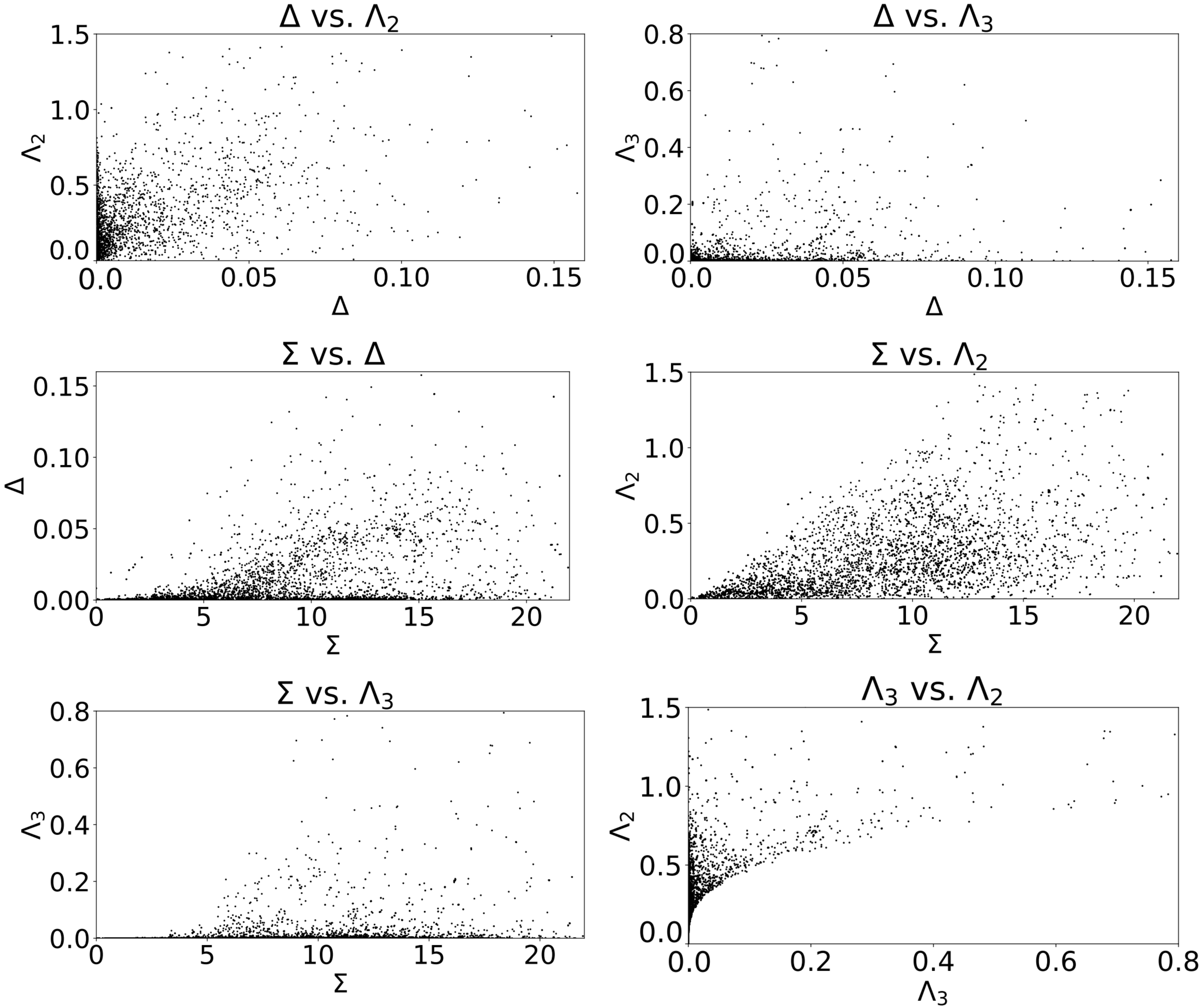}
      \caption{Pairwise plots of the $\Delta$, $\Sigma$, $\Lambda_{2}$, and $\Lambda_{3}$ descriptors computed using the dataset described in Sec.~\ref{sec:level3} by the Pyrovskite package.}
      \label{fig:plot}
   \end{centering}
\end{figure}
The Pyrovskite package can be used to automatically detect the B-site and X-site ions, identify the corresponding octahedra, and compute the $\Delta$, $\Sigma$, $\Lambda_2$, and $\Lambda_3$ descriptors. Figure~\ref{fig:plot} and Table~\ref{tab:table1} show the correlation of the above four parameters computed for a diverse dataset comprising over 1000 perovskites and thereby over 5000 perovskite octahedra. The full details of this dataset, as well as example code for the reproduction of such analysis can be found in the Supplemental Material \cite{supp2022mat} Sec.~\ref{sec:level1}-~\ref{sec:level3}. The correlation coefficients between the $\Lambda_2$ descriptor and the $\Delta$ and $\Sigma$ descriptors are the highest at $0.57$, as is to be expected as the in-plane diagonal displacement necessitates an increased deviation from average bond lengths and angles. Nonetheless, the descriptors still maintain a substantial degree of noncollinearity. Additionally, the correlation between $\Lambda_3$ and the $\Delta$ and $\Sigma$ descriptors is quite low at $0.36$, and $0.21$ respectively. The large difference in correlation of the new $\Lambda_2$ and $\Lambda_3$ parameters with the previously established $\Delta$ and $\Sigma$ descriptors offers a convenient way to select for a feature which adequately quantifies the diagonal B-cation displacement, while avoiding collinearity with another feature describing the octahedral distortions present in the system.

\begin{table}
\caption{\label{tab:table1}Pearson correlation coefficients of the $\Sigma$, $\Delta$, $\Lambda_2$, and $\Lambda_3$ descriptors across the dataset used in the present study.}
\begin{ruledtabular}
\begin{tabular}{ccccc}
Correlation & $\Sigma (^\circ)$ & $\Delta (\AA)$ & $\Lambda_2 (\AA)$ & $\Lambda_3 (\AA^3)$\\
 Coefficients &&&&\\
\hline
$\Sigma$ &1.00&0.41&0.57&0.21\\
$\Delta$ &0.41&1.00&0.57&0.36 \\
$\Lambda_2$&0.57&0.57&1.00&0.48 \\
$\Lambda_2$&0.21&0.36&0.48&1.00\\
\end{tabular}
\end{ruledtabular}
\end{table}

\subsection{\label{sec:level32}Procedural generation of perovskite crystal structures}

\begin{figure}[!ht]
   \begin{centering}
    \includegraphics[width=0.9\columnwidth]{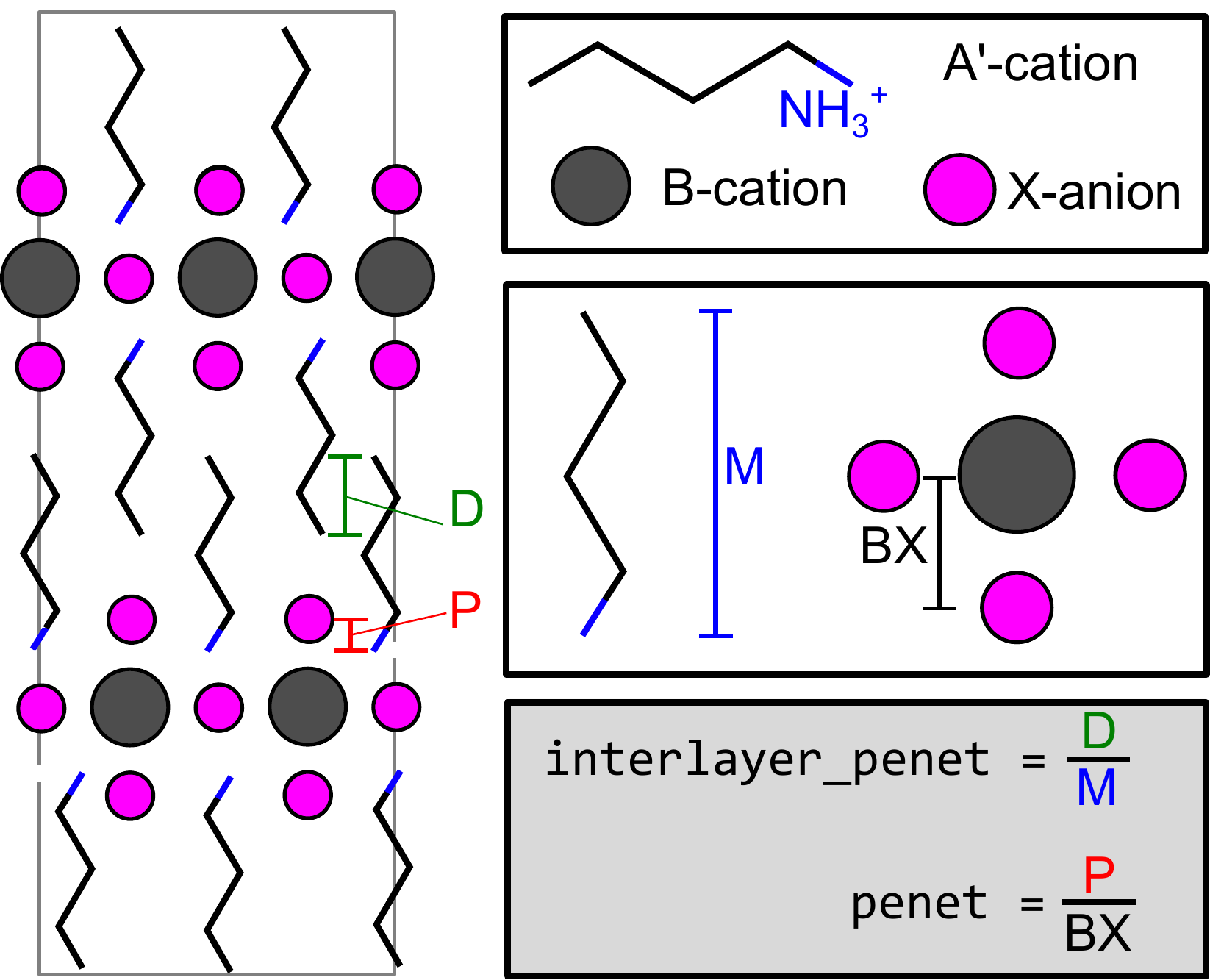}
      \caption{Schematic representation detailing the procedural generation of 2DPKs by the Pyrovskite builder module. The $\textsf{interlayer\_penet}$ parameter describes the degree of interlocking present between A’ cations in the 2DRP phase, and the $\textsf{penet}$ parameter describes the penetration depth of the A’ cation into the adjacent inorganic octahedra. Hydrogens of the A'-cation are omitted for clarity, but we note they are included for the computation of the molecule length $M$.}
      \label{fig:2dpk}
   \end{centering}
\end{figure}

The Pyrovskite package contains the builder module which allows for the rapid generation of both 2D and 3D perovskites in multiple phases, while including organic A-site cations and A'site cations where applicable. The builder module leverages the atomic simulation environment (\textsf{ASE}) Atoms object in order to facilitate construction of the perovskite systems \cite{larsen2017atomic}. For purely inorganic, bulk 3D perovskites, multiple tools exist for the generation of crystal structures of varying components such as \textsf{pymatgen} \cite{ong2013python}. However, for the incorporation of 2DPKs, and organic A-site cations, the authors are not aware of any software packages which address this need. The Pyrovskite builder module utilizes functionality of the \textsf{ASE} Atoms object for the generation of all perovskite structures. A-site cations can be passed as either an Atoms object, or as a path to one of various input structure files such as .xyz, .sdf, .pdb, among others. A-site cation orientation can be adjusted, and the center of mass is situated at the corresponding crystal site of the given perovskite phase. In the case of 2DPKs, Fig.~\ref{fig:2dpk} demonstrates the key factors which determine the crystal structure. For example, a 2DPK in the $n=2$, 2DRP phase can be created by simply by specifying the path to any organic cations, the elemental symbol of any inorganic ions, the layer thickness $n$, and the desired $B-X$ bond length:\\

\begin{python}
from pyrovskite.builder import make_2drp
A = "./methylammonium.xyz"
Ap = "./butylammonium.xyz"
org_A_2drp = make_2drp(Ap, A, "Pb", "I", 2, 3)
in_A_2drp = make_2drp(Ap, "Cs", "Pb", "I", 2, 3)
\end{python}

A’ spacing cations locations are determined by the perovskite phase, as well as by the degree to which they penetrate into the inorganic BX$_{6}$ octahedra. Additionally, in the case of 2DRP phases, A’ cation placement depends on the degree to which the adjacent organic A’ cations interlock with one another. Sensible defaults are chosen for the interlayer penetration, $\textsf{interlayer\_penet}$, and the octahedral penetration, $\textsf{penet}$ of $0.4$ and $0.3$ respectively. As shown in Fig.~\ref{fig:2dpk}, $\textsf{penet}$ is given in terms of a fraction of the B-X bond length for the system, and $\textsf{interlayer\_penet}$ is given in terms of a fraction of the A’ cation length. The builder module allows for structure generation requiring only specification of the elemental and molecular building components, the perovskite phase, and the layer thickness in the case of 2DPKs. Additionally, more fine-grained control is available to the user such as the ability to 1) modify the orientation of all organic A and A’ cations, 2) adjust the layer penetration, and 3) introduce disorder to the created structure. The builder module is capable of creating 1) 3D ABX$_{3}$ perovskites, 2) 3D A$_{2}$BB’X$_{6}$ double perovskites, 3) 2DRP perovskites, 4) DJ perovskites, 5) 2D monolayer perovskites, 6) 2DRP double perovskites, 7) DJ double perovskites, and 8) 2D monolayer double perovskites. The package comes with numerous organic A-site, and A’ spacing cations, and including additional organic cations is straightforward and described in the documentation. Structures created using the builder module can be saved as any typical crystal structure file compatible with the \textsf{pymatgen} or \textsf{ASE} packages. Additionally, the builder module can supply the user with input files for subsequent electronic structure calculations using either GPAW or xTB via CP2K \cite{enkovaara2010electronic, bannwarth2019gfn2, hutter2014cp2k, kuhne2020cp2k}.

\section{\label{sec:level4}Capabilities}
\begin{figure}[!ht]
   \begin{centering}
    \includegraphics[width=0.9\columnwidth]{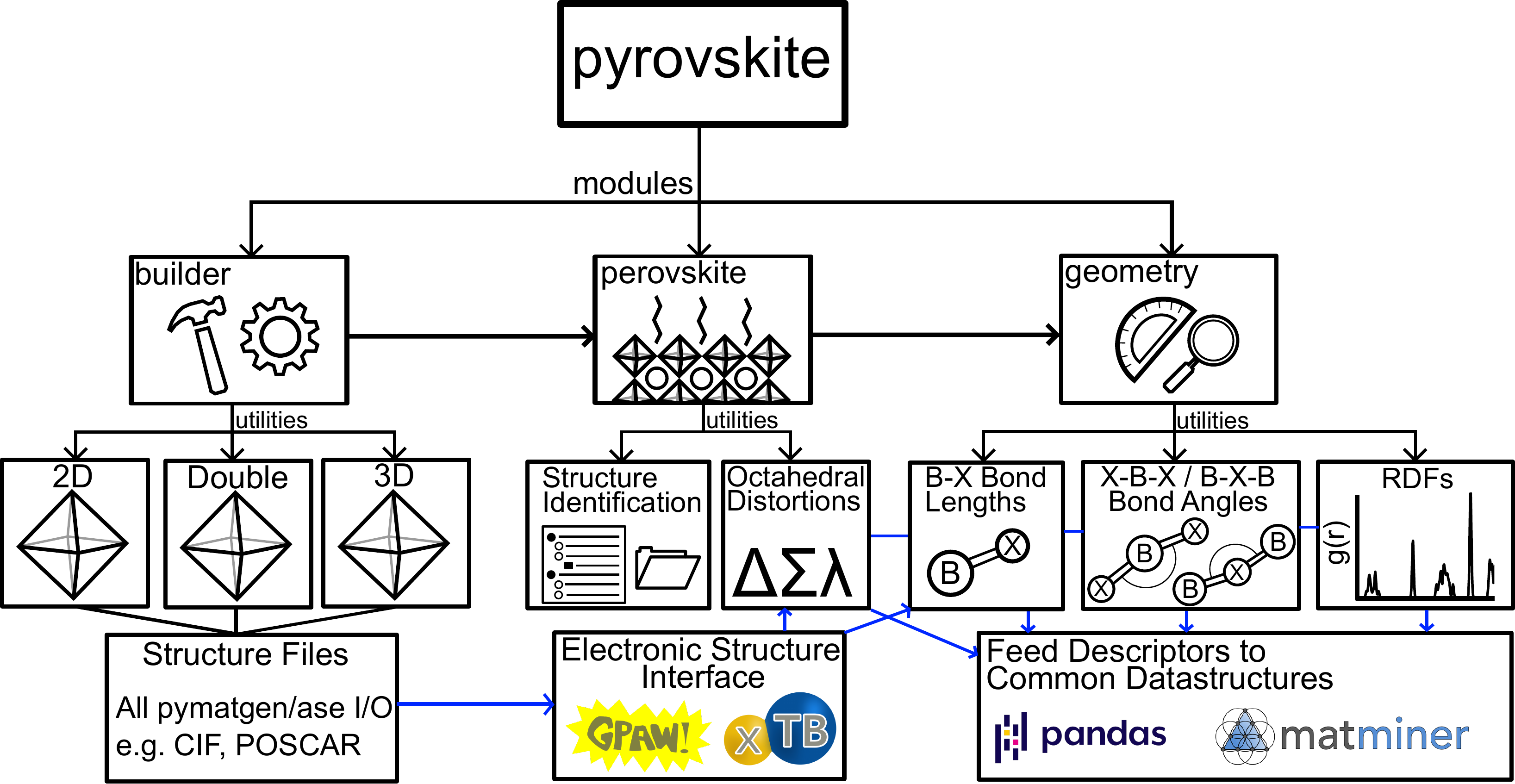}
      \caption{Schematic representation for the structure of the Pyrovskite codebase. A typical workflow beginning from structure generation, subsequent electronic structure calculations, computation of geometric properties, and exporting of data for further analysis is highlighted in blue.}
      \label{fig:flowchart}
   \end{centering}
\end{figure}

The Pyrovskite codebase contains numerous utilities to aid in both high-throughput and fine-grained computational workflows involving perovskite systems, including crystal structure creation, electronic structure input file generation, structural analysis, and interfacing with commonly used tools for subsequent data analysis or machine learning. The central datatype of the package is the Perovskite class from which most functionality of the geometry and builder modules is exposed directly [as schematically shown in Fig.~\ref{fig:flowchart}]. The Perovskite class can be instanced either through the creation of a structure in the builder module or by passing any perovskite crystal structure by file name or as an $\textsf{ASE}$ Atoms object. If instanced using a crystal structure, additional information can be passed including the identity of the A, A’, B, and X building components but is not necessary. The class is robust to the automatic detection of B-site cations and X-site anions, even in the presence of organic molecules which contain common X-site anions, such as fluorinated spacer molecules. Furthermore, the class detects B’ cations in the case of double perovskites of either three or two dimensions. This greatly aids in applications which parse large datasets using tools like the Materials Project API where such information may not be directly available \cite{jain2013commentary}. The automatic identification of all unique octahedra for computation of the distortion parameters discussed in Sec.~\ref{sec:level2} and~\ref{sec:level3} is also robust to a broad range of perovskite phases and chemical compositions, making the codebase suitable for most researchers in the field. Much of the functionality of the Pyrovskite package is also applicable to other systems which host octahedral coordination environments, including molecular systems containing transition-metal complexes as well as condensed matter systems like metal-organic frameworks.\\

A common use case for the package in structure generation has already been detailed in Sec.~\ref{sec:level3}, but here we highlight a more complete use case beginning with structure preparation and writing an xTB input file for usage with CP2K:\\

\begin{python}
from pyrovskite.builder import make_2drp
from pyrovskite.perovskite import Perovskite
A = "./methylammonium.xyz"
Ap = "./butylammonium.xyz"
new_perov = make_2drp(Ap, A, "Pb", "I", 2, 3.1)
perovskite_2drp = Perovskite(new_perov)
perovskite_2drp.write_xTB("perovskite_2drp.inp")
\end{python}

At this point, the user can run the electronic structure calculation which in the above scenario defaults to a geometry optimization. The subsequent optimized structure can then be read by the Pyrovskite package to compute octahedral distortions, visualize the isolated octahedra, and create publication quality figures of the B-X bond length, X-B-X bond angle, and B-X-B bond angle distributions, as well as partial radial distribution functions of the resulting system:\\

\begin{python}
# Import and load in the optimized perovskite
from pyrovskite.perovskite import Perovskite
opt  = Perovskite("./xTB_opt_perov_2drp.cif")

# Compute octahedral distortion parameters.
delta = opt_perovskite.compute_delta()
sigma = opt_perovskite.compute_sigma()
lambda3, lambda2 = \
opt_perovskite.compute_lambda()

# Visualize the identified octahedra.
opt_perovskite.plot_vertices()

# Plot bond, angle distributions, and pRDFs
opt_perovskite.plot_distances()
opt_perovskite.plot_angles()
opt_perovskite.plot_rdf()
\end{python}

These features also work seemlessly with double perovskite systems where B’-X bonds, as well as X-B’-X, B’-X-B, and B’-X-B’ angles will be included by default where applicable if a double perovskite structure is provided. All computed properties are stored in the Perovskite object, and in-built functions for exporting structures to \textsf{pandas} DataFrames exist for easy integration into, for example, \textsf{MatMiner} workflows.\\

\section{\label{sec:level5}Conclusion and Outlook}

Our open-source package Pyrovskite provides an all-inclusive platform for the generation and analysis of perovskite systems, as well as a seamless interface with multiple electronic structure codes and external resources to aid in large-scale data analysis. Additionally, the octahedral distortion descriptors introduced here generalize to any octahedral coordination environment, greatly expanding the utility of the package. The octahedra of both molecular systems, as well as non-perovskite condensed matter systems can be automatically detected, allowing for the facile computation of octahedral distortion parameters beyond the scope of perovskites. The codebase is under active development and will in the future contain automatic phase detection, characterization of octahedral tilt patterns, and pre-trained models for the prediction of material properties from both chemical formulas, and crystal structures.

\section*{Data Availability}

This software package is available at \href{http://github.com/r2stanton/pyrovskite}{Pyrovskite}.

\nocite{*}

\begin{acknowledgments}
This work is supported by the U. S. National Science Foundation Engineering Research Initiation Award ECCS-2138728.
\end{acknowledgments}

\bibliography{StantonTrivedi-Pyrovskites}
 
\end{document}